\documentclass[a4paper,11pt]{article}
\usepackage{pos}
\usepackage{ulem}
\usepackage{amsmath}

\title{Extraction of the $S$-wave and $P$-wave $DD^*$ scattering phase shifts using twisted boundary conditions}
\ShortTitle{Extraction of the $S$-wave and $P$-wave $DD^*$ scattering phase shifts .....}

\author*[a]{Masato Nagatsuka}
\author[a]{Shoichi Sasaki}

\affiliation[a]{Department of Physics, Tohoku University,\\
Sendai 980-8578, Japan}

\emailAdd{masato.nagatsuka.r4@dc.tohoku.ac.jp}
\emailAdd{ssasaki@nucl.phys.tohoku.ac.jp}

\abstract{
We present results of a lattice study of the $S$-wave and $P$-wave $DD^*$ scattering phase shifts using Lüscher’s method under the twisted boundary conditions to investigate the doubly charmed tetraquark $T_{cc}^+$ observed by the LHCb collaboration.  Although the scattering phase shift at zero momentum gives information about the number of bound states according to Levinson’s theorem, Lüscher’s method under the periodic boundary condition only accesses the scattering phase shifts at some discrete momenta and is not suitable for watching the signal of bound state formation.
On the other hand, the twisted boundary condition has the advantage that the scattering phase shift at any momentum can be calculated and that not only the $S$-wave scattering phase shift but also the $P$-wave scattering phase shift can be obtained simultaneously.
In this study, we perform the simulation for the $DD^*$ and $BB^*$ systems in the $I=0$ channel using 2+1 flavor PACS-CS gauge ensembles simulated at $m_\pi=295$ and 411 $\mathrm{MeV}$.
}

\FullConference{The 41st International Symposium on Lattice Field Theory (LATTICE2024)\\
 28 July - 3 August 2024\\
Liverpool, UK\\}

\begin{document}
\maketitle

%%%%%%
\section{Introduction}
\label{sec:intro}
The LHCb collaboration observed a new resonance interpreted to be the doubly charmed tetraquark state $T_{cc}^+$ in 2021 \cite{LHCb:2021vvq,LHCb:2021auc}. It is considered to have isospin $I=0$ and quark content $cc\bar u\bar d$ by the experiment. In addition to the discovery, 
some model calculations indicate that 
doubly-heavy tetraquark system
such as $QQ\bar u\bar d$ has a bound state if the mass of quark $Q$ is sufficiently heavy, comparable to that of the bottom quark~\cite{Manohar:1992nd,Dai:2022ulk}.

In order to theoretically confirm the signal of the doubly charmed tetraquark, we calculate the scattering phase shift between $D$ and $D^*$ mesons with lattice QCD using L\"uscher's method under twisted boundary conditions \cite{Luscher:1990ux, Ozaki:2012ce}. There are several calculation of the scattering phase shift using L\"uscher's method~\cite{Padmanath:2022cvl, Collins:2024sfi, Chen:2022vpo} and using the HAL QCD method~\cite{Lyu:2023xro}. Although these authors aim to find the $S$-matrix pole located below the $DD^*$ threshold from the $DD^*$ scattering phase shift, the situation is complicated by the fact that the left-hand cut of $DD\pi$ just below the $DD^*$ threshold could give rise to the singularity to the $DD^*$ scattering amplitude~\cite{Du:2023hlu}.
We therefore do not aim to determine the pole condition, rather 
focus on the detailed information on the scattering phase shift just above the $DD^*$ threshold in this study.

In general, Levinson's theorem tells us that the number of bound states $n$ and the $l$-wave scattering phase shift at zero momentum $\delta_l(0)$ are related as $\delta_l(0)=n\pi$. Therefore, we can predict the number of bound states in the system by tracing the behavior of scattering phase shift around zero momentum~\cite{Sasaki:2006jn}.
However, L\"uscher's method under the periodic boundary condition only gives the scattering phase shift at a few discrete momenta because the momentum of the center of mass $\pmb{P}$ %$\pmb{P}$ 
is discretized as $(\frac{L\pmb{P}}{2\pi})=0,~1,~2,\cdots$ due to the finiteness of the volume.

In this context, twisted boundary conditions plays an important role. We impose the following boundary condition to the wave function $\Psi(\pmb{x})$ as
\begin{equation}
\Psi(\pmb{x}+L\pmb{e}_k)=e^{i\theta_k}\Psi(\pmb{x}),
\end{equation}
where $\pmb{e}_k$ are unit vectors in the direction of the $k$ axis ($k=x,y,z$) and $\theta_k$  are real valued twisted angles. Under the twisted boundary condition, the momentum of a plane wave $\pmb{p}$ are discretized as
\begin{equation}
\pmb{p} = \frac{2\pi\pmb{n}}{L}+\boldsymbol{\theta},\quad\pmb{n}\in\mathbb{Z}^3
\end{equation}
with a twist angle vector $\pmb{\theta}=(\theta_x, \theta_y, \theta_z)$ for the free case.
By adjusting the arbitrary parameter $\theta_k$, we can change the ground state energy of two-hadron scattering system, but this requires modifications to the original L\"uscher's method as explained in Sec.~\ref{sec:TheorFram}. We can make use of this formula to calculate both $S$-wave and $P$-wave scattering phase shifts at low energies in detail.

We therefore perform the calculation of scattering phase shift for the  both $DD^*$ and $BB^*$ systems using L\"uscher's method under twisted boundary conditions to explore the behavior of the low-energy scattering near the $DD^*$ and $BB^*$ threshold. 

%%%%%%
\section{Formulation}
\label{sec:TheorFram}
\subsection{L\"uscher's finite size formula and the calculation strategy}
For the case of $S$-wave ($l=0$) scattering phase shift $\delta_0(k)$, it is possible to be calculated from energy spectra using L\"uscher's formula under
the periodic boundary condition
\begin{equation}
\cot\delta_0(k)=\frac{1}{\pi^{3/2}q}Z_{00}(1;q^2),\qquad q=\frac{Lk}{2\pi}, \label{eq:Luscher_formula_origin}
\end{equation}
where the higher partial-wave ($l\ge4$) contributions are ignored.
Since L\"uscher proposed the equation, we have seen many different kinds of extensions.
Eventually, if the twisted boundary conditions are employed and the higher partial-wave ($l\ge2$) contributions are ignored, the formula can be summarized 
as
\begin{equation}
\left|\begin{matrix}
  \cot\delta_0(k)-\mathcal{M}_{{\rm SS}}^{\boldsymbol{\theta}}(q) & \mathcal{M}_{{\rm SP}}^{\boldsymbol{\theta}}(q) \\
  \mathcal{M}_{{\rm SP}}^{\boldsymbol{\theta}}(q)^* & \cot\delta_1(k)-\mathcal{M}_{{\rm PP}}^{\boldsymbol{\theta}}(q)  
\end{matrix}\right|=0,
\label{eq:Luscher_formula_twist}
\end{equation}
where $\mathcal{M}_{{\rm SS}}^{\boldsymbol{\theta}}(q),~\mathcal{M}_{{\rm SP}}^{\boldsymbol{\theta}}(q)$ and $\mathcal{M}_{{\rm PP}}^{\boldsymbol{\theta}}(q)$ depend on the direction of $\boldsymbol{\theta}$, as shown in Table \ref{Tab:MatrixElem}~\cite{Ozaki:2012ce}. $w_{lm}$ that appears in the table are defined as
\begin{equation}
w_{lm}=\frac{1}{\pi^{3/2}\sqrt{2l+1}q^{l+1}}Z^{{\pmb \theta}}_{lm}(1;q^2)^*.
\label{eq:def_wlm}
\end{equation}

%
%\begin{table}[t]
\begin{table*}
\begin{center}
%\begin{ruledtabular}
\begin{tabular}{cccccccc}
\hline
Twist angle & $(0,0,\theta)$ & $(\theta,\theta,0)$ & $(\theta,\theta,\theta)$ \\
Symmetry & $C_{4v}$ & $C_{2v}$ & $C_{3v}$ \\
label & $[001]$ & $[110]$ & $[111]$ \\
\hline
$\mathcal{M}_{{\rm SS}}^{\boldsymbol{\theta}}(q)$ & $w_{00}$ & $w_{00}$ & $w_{00}$ \\
$\mathcal{M}_{{\rm SP}}^{\boldsymbol{\theta}}(q)$ & $i\sqrt{3}w_{10}$ & $i\sqrt{6}w_{11}$ & $i3w_{10}$ \\
$\mathcal{M}_{{\rm PP}}^{\boldsymbol{\theta}}(q)$ & $w_{00}+2w_{20}$ & $w_{00}-w_{20}-i\sqrt{6}w_{22}$ & $w_{00}-i2\sqrt{6}w_{22}$ \\
\hline
\end{tabular}
\caption{Definitions of $\mathcal{M}_{{\rm SS}}^{\boldsymbol{\theta}}(q),~\mathcal{M}_{{\rm SP}}^{\boldsymbol{\theta}}(q)$ and $\mathcal{M}_{{\rm PP}}^{\boldsymbol{\theta}}(q)$ for each twist angle ($0< |\theta| < \pi$).}
\label{Tab:MatrixElem}
%\end{ruledtabular}
\end{center}
\end{table*}

The generalized zeta function is also defined as
\begin{equation}
Z^{{\pmb \theta}}_{lm}(1;q^2)=\sum_{\pmb{r}\in\Gamma_{\boldsymbol{\theta}}}\frac{\mathcal{Y}_{lm}(\pmb{r})}{(\pmb{r}^2-q^2)},
\end{equation}
where $\Gamma_{\boldsymbol{\theta}}=\{\pmb{r}|\pmb{r}=\pmb{n}+\frac{\boldsymbol{\theta}}{2\pi},\pmb{n}\in\mathbb{Z}^3\}$ and $\mathcal{Y}_{lm}(\pmb{r}) =|\pmb{r}|^l Y_{lm}(\pmb{r})$.

This formula claims that the angular momentum states can be highly mixed each other and it is not reasonable to truncate the effect of 
$P$-wave ($l=1$) scattering for $0 < |\theta_k| < \pi$. In our research, we make use of the property to extract not only the $S$-wave scattering phase shift, but also the $P$-wave scattering phase shift according to the following strategy divided into three steps \cite{Ozaki:2012ce}.

\begin{enumerate}
%%%%% Step 1 %%%%%
\item Calculate the $S$-wave scattering phase shifts using Eq.~\eqref{eq:Luscher_formula_origin} for special angles $\boldsymbol{\theta} =(0,0,0)$, $(0,0,\pi)$, $(\pi,\pi,0)$ and $(\pi,\pi,\pi)$.
Then, we interpolate the data using the effective range expansion
\begin{equation}
k\cot\delta_0(k)=\frac{1}{a_0}+\frac{1}{2}r_0k^2+v_0k^4.
\label{eq:ERE_Swave}
\end{equation}
This treatment is valid near the threshold, where the higher partial-wave ($l\ge 2$) contributions are safely ignored.
Indeed, there are no mixing between $S$-wave and $P$-wave states for these special angles since the corresponding point group maintains a center of inversion symmetry.

%%%%% Step 2 %%%%%
\item Calculate the $P$-wave scattering phase shift according to  Eq.~\eqref{eq:Luscher_formula_twist} for $\boldsymbol{\theta}=(\theta,\theta,\theta)$ expressed as
\begin{equation}
\cot\delta_1(k)=w_{00}(q)+2\sqrt{6}{\rm{Im}}\{w_{22}(q)\}+\frac{9w_{10}^2(q)}{\cot\delta_0(k)-w_{00}(q)}.
\label{eq:Luscher_formula_twist3d}
\end{equation}
We use the interpolated data sets of $\cot\delta_0$ given by  Eq.~\eqref{eq:ERE_Swave} as inputs in the right hand side of Eq.~\eqref{eq:Luscher_formula_twist3d}. We also interpolate the obtained $P$-wave scattering phase shift according to the effective range expansion up to $\mathcal{O}(k^2)$.

%%%%% Step 3 %%%%%
\item Finally calculate the $S$-wave scattering phase shift near the threshold using Eq.~\eqref{eq:Luscher_formula_twist} for $\boldsymbol{\theta}=(0,0,\theta)$ expressed as
\begin{equation}
\cot\delta_0(k)=w_{00}(q)+\frac{3w_{10}^2(q)}{\cot\delta_1(k)-w_{00}-2w_{22}(q)}
\label{eq:Luscher_formula_twist1d}
\end{equation}
with help of the interpolated data sets of $\cot \delta_1$ obtained from the second step.

\end{enumerate}

There are two advantages in this strategy. First, 
$S$-wave scattering phase shifts near the threshold can be obtained with high resolution.
Second, it is also possible to simultaneously obtain the $P$-wave scattering phase shift only by calculating the irreducible representation (irrep) $A_1$ of two-hadron states.

\subsection{Calculation under the twisted boundary condition}
In our simulation we use the twisted boundary condition for the charm and bottom quarks while the periodic boundary condition is used for light quarks. We calculate the charm and bottom quark propagators under the twisted boundary conditions as follows.

Suppose quark fields $q_{\boldsymbol{\theta}}(\pmb{x},t)$ satisfy the twisted boundary condition
\begin{equation}
q_{\boldsymbol{\theta}}(\pmb{x}+L \pmb{e}_j,t)=e^{i\theta_j}q_{\boldsymbol{\theta}}(\pmb{x},t).
\end{equation}
We can introduce new fields
\begin{equation}
q'(\pmb{x},t)=e^{-i\boldsymbol{\theta} \cdot\pmb{x}/L} q_{\boldsymbol{\theta}}(\pmb{x},t),
\end{equation}
which obey to the periodic boundary condition. Therefore, for the 
Wilson-type fermions, we can calculate quark propagators subject to the twisted boundary condition with a simple modification $U_{x,\mu}\to e^{i\theta_\mu a/L}U_{x,\mu}$ to gauge configurations $U_{x,\mu}$ where $\theta_\mu=(\boldsymbol{\theta},0)$ and the lattice spacing $a$.

In our calculations, we use wall-source operators so that the two-hadron interpolating operators are automatically projected to the trivial irreducible representations of the point group.
The wall source operator for $q'(\pmb{x},t)$ can be rewritten in terms of $q_{\boldsymbol{\theta}}(\pmb{x},t)$ as
\begin{equation}
q(\pmb{p},t)=\sum_{\pmb{x}}q'(\pmb{x},t)=\sum_{\pmb{x}}q_{\boldsymbol{\theta}}(\pmb{x},t)e^{i\boldsymbol{\theta} \cdot\pmb{x}/L},
\end{equation}
where $\pmb{p}=\boldsymbol{\theta}/L$. Thus, the wall-source quark operator with twisted angle $\boldsymbol{\theta}$ can be considered as the momentum $\pmb{p}$ projected operator. We then use wall-source operators to construct the hadron ($h$) interpolating operator at the source as
\begin{equation}
O^{W,k}_h(\pmb{p},t_{\mathrm{src}})= \sum_{\pmb{x}}\bar q_f(\pmb{x},t_{\mathrm{src}}) \Gamma_{k}\sum_{\pmb{y}}q_{f'}(\pmb{y},t_{\mathrm{src}})
\end{equation}
with $\pmb{p}=(\boldsymbol{\theta}_f-\boldsymbol{\theta}_{f'})/L$, 
while we use local operators
to construct the hadron ($h$) interpolating operator at the sink
\begin{equation}
O^{L,k}_h(\pmb{x},t_{\mathrm{snk}})=\bar q_f(\pmb{x},t_{\mathrm{snk}}) \Gamma_k q_{f'}(\pmb{x},t_{\mathrm{snk}}),
\end{equation}
where $f$ and $f'$ denote flavor indices and
$\Gamma_k$ is a gamma matrix. $\Gamma_k=\gamma_k~(k=1,2,3)$ is chosen for spin-1 mesons, and $\Gamma_k=\gamma_5$~($k=5$) is chosen for spin-0 mesons. 
The index $k$ is omitted hereafter. In case of $h=D$ and $D^*$ ($B$ and $B^*$) mesons, the twist angles of up and down quarks are fixed to be $\boldsymbol{\theta}_f=\pmb{0}$, while the twist angle of charm (bottom) quarks is varied so that the momentum $\pmb{p}$ is finite. 

In this study, we only consider the center-of-mass of the two-hadron system. For the two-hadron ($DD^*$) interpolating operator at the source, we use wall operators given as
\begin{align}
Q^W_{DD^*}(\pmb{P}=\pmb{0},\pmb{p},t_{\mathrm{src}})=O^{W}_{D}(\pmb{p},t_{\mathrm{src}})O^{W}_{D^*}(-\pmb{p},t_{\mathrm{src}}),
\end{align}
where the $D$ and $D^*$ operators carry the opposite momentum so as to make zero total momentum of the $DD^*$ system ($\pmb{P}=\pmb{0}$). Non-zero $\pmb{p}$ is responsible for
non-zero relative momentum between the $D$ and $D^*$ mesons.
For the sink operator, we use a simple product of 
two local operators and take summation over the spatial sites independently for each operator as
\begin{equation}
Q^L_{DD^{*}}(\pmb{P}=\pmb{0},t_{\mathrm{snk}})=\sum_{\pmb{x}}
O^{L}_{D}(\pmb{x},t_{\mathrm{snk}})
\sum_{\pmb{y}}O^{L}_{D^*}(\pmb{y},t_{\mathrm{snk}}),
\end{equation}
where the total momentum of the $DD^*$ system is projected onto zero 
momentum, when the twist angles of the charm quarks involved in $D$ and $D^{*}$ mesons are taken in opposite directions with the same angle size.

The usage of the wall-source operators indicates that 
the resulting two-hadron correlators are already projected to the trivial irrep $A_1$ of any point group as discussed in Ref.~\cite{Ozaki:2012ce}.
In addition, for the case of the $DD^*$ or $BB^*$ scattering, since there is only a single spin state, no spin projection is required.
For the isospin $I$ of the $DD^*$ or $BB^*$ systems, we focus only on the $I=0$ channel in this talk.

\section{Numerical Results}
\label{sec:NumResults}
\subsection{Simulation Details}
We apply L\"uscher's method to explore the $DD^*$ and $BB^*$ scatterings at low energies.
For this purpose we perform lattice QCD simulations on a lattice $L^3\times T = 32^3\times 64$ with 2+1 flavor PACS-CS gauge configurations, where the simulated pion masses are $m_\pi=295$ and 411 $\mathrm{MeV}$. 
Simulation parameters of PACS-CS gauge configurations are summarized in Table~\ref{Tab:LatParam}.

We use nonperturbatively improved clover fermions for up and down quarks while relativistic heavy quark (RHQ) action is used for charm and bottom quarks. Parameters of clover fermions and the RHQ action used in this work are listed in Table~\ref{Tab:QuarkParam}. The RHQ parameters defined in a Tsukuba-type action~\cite{Aoki:2001ra,Kayaba:2006cg} have been properly determined in Ref.~\cite{Kawanai:2011jt}.

%
%\begin{table}[t]
\begin{table*}
\begin{center}
%\begin{ruledtabular}
\begin{tabular}{cccccccc}
\hline
$\beta$ \ & \ $a$ (fm) \ &  $L^{3} \times T$ & \ $\sim La$ (fm) & $c_{\rm SW}$ \\
\hline
$1.9$  \ & $0.0907(13)$ \ & $32^{3} \times 64$ \ & $2.9$ \ &  $1.715$  \\
\hline
\end{tabular}
\caption{Simulation parameters of $2+1$ flavor PACS-CS gauge configurations, generated using the Iwasaki gauge action and Wilson clover fermions~\cite{Aoki:2008sm}.}
\label{Tab:LatParam}
%\end{ruledtabular}
\end{center}
\end{table*}

\subsection{Analysis of scattering phase shifts}
Making use of the operators defined in Sec.~\ref{sec:TheorFram}, we calculate masses of $D,~D^*,~B$ and $B^*$ mesons and energies of
the $DD^*$ and $BB^*$ systems. The masses of each mesons are tabulated in  Table.~\ref{Tab:Spect}. We are also able to calculate the scattering momentum $k$ from the total energy of two-meson system by using the relation 
\begin{equation}
W=\sqrt{k^2+M_h^2}+\sqrt{k^2+M_{h^*}^2}\quad(h=D,~B).
\end{equation}
After we obtain $k^2$ for each twisted angle, we can use L\"uscher's method to calculate the $S$-wave scattering phase shift $\delta_0(k)$ and $P$-wave scattering phase shift $\delta_1(k)$ according to the three steps described in Sec.~\ref{sec:TheorFram}.

\begin{figure*}
\centering
\includegraphics*[width=.48\textwidth,bb=0 0 792 612,clip]{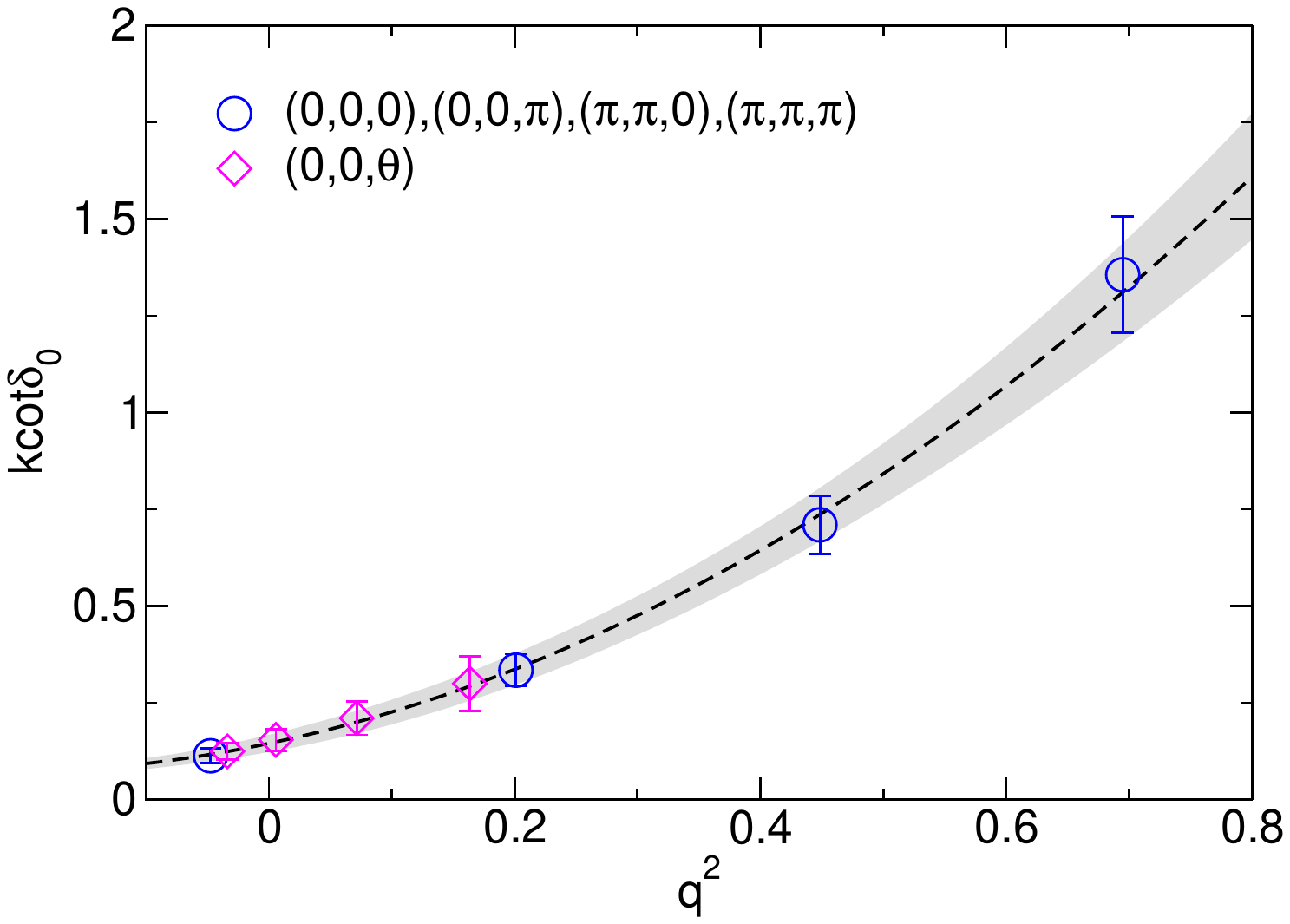} 
\includegraphics*[width=.48\textwidth,bb=0 0 792 612,clip]{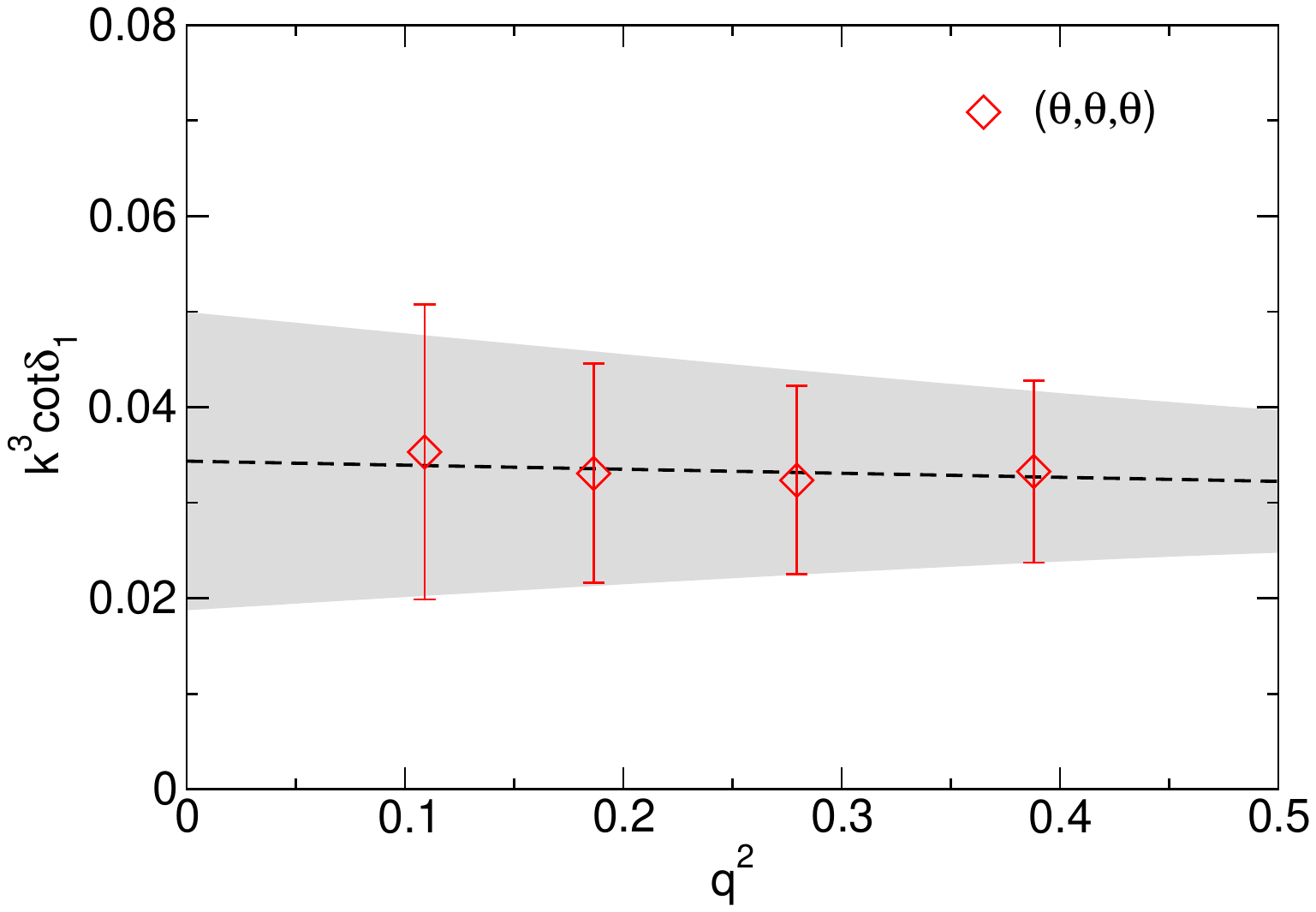}
\caption{
{\bf Left}: $k\cot\delta_0$ as a function of $q^2$ for the case of the $I=0$ $DD^*$ scattering at $m_\pi=411~\mathrm{MeV}$. 
{\bf Right}: $k^3\cot\delta_1$ as a function of $q^2$ for the case of the $I=0$ $DD^*$ scattering at $m_\pi=411~\mathrm{MeV}$.
} 
\label{fig:plot_kcot}
\end{figure*}

First, we calculate $S$-wave scattering phase shift $\delta_0(k)$ for specific four angles
$\boldsymbol{\theta} =(0,0,0)$, $(0,0,\pi)$, $(\pi,\pi,0)$ and $(\pi,\pi,\pi)$
where we can use Eq.~\eqref{eq:Luscher_formula_origin}. As a typical example, we show the $k\cot\delta_0$ as a function of $q^2$ for the case of the $I=0$ $DD^*$ scattering with $m_\pi=411~\mathrm{MeV}$ in the left panel of Fig~\ref{fig:plot_kcot}. We observe that $k \cot\delta_0$ is monotonically increasing within this range and we interpolate the four data points expressed as circles.
Next, we calculate $P$-wave scattering phase shift 
$\delta_1(k)$ from data sets of $\pmb{\theta}=(\theta,\theta,\theta)$ using L\"uscsher's formula~\eqref{eq:Luscher_formula_twist3d}
with a help of information obtained beforehand on $S$-wave. We observe the behavior of $k^3\cot\delta_1$ as a function of $q^2$ in the right panel of Fig.~\ref{fig:plot_kcot}. We fit these data as the linear function of $q^2$.
We then can calculate the $S$-wave scattering phase shifts near the threshold from  
data sets of $\pmb{\theta}=(0,0,\theta)$ using L\"uscsher's formula~\eqref{eq:Luscher_formula_twist1d}
with a help of the $P$-wave information obtained from the data sets of $\pmb{\theta}=(\theta,\theta,\theta)$.
As shown in the left panel of Fig~\ref{fig:plot_kcot}, 
the results of $S$-wave $k\cot\delta_0(k)$ obtained near the threshold (diamond symbols) fill a gap between two data points obtained from $\boldsymbol{\theta} =(0,0,0)$ and $(0,0,\pi)$.

\begin{figure*}
\centering
\includegraphics*[width=.48\textwidth,bb=0 0 792 612,clip]{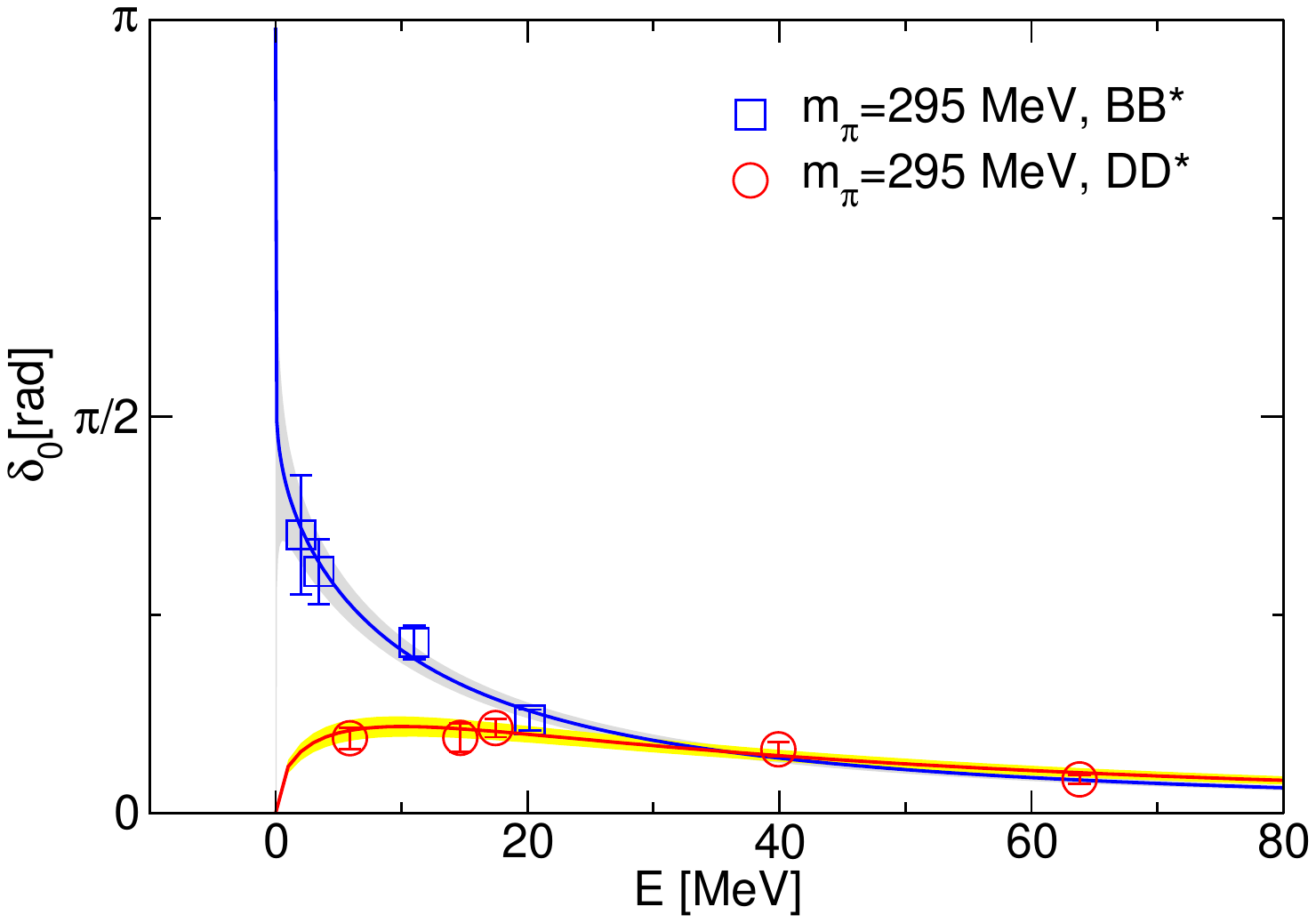} 
\includegraphics*[width=.48\textwidth,bb=0 0 792 612,clip]{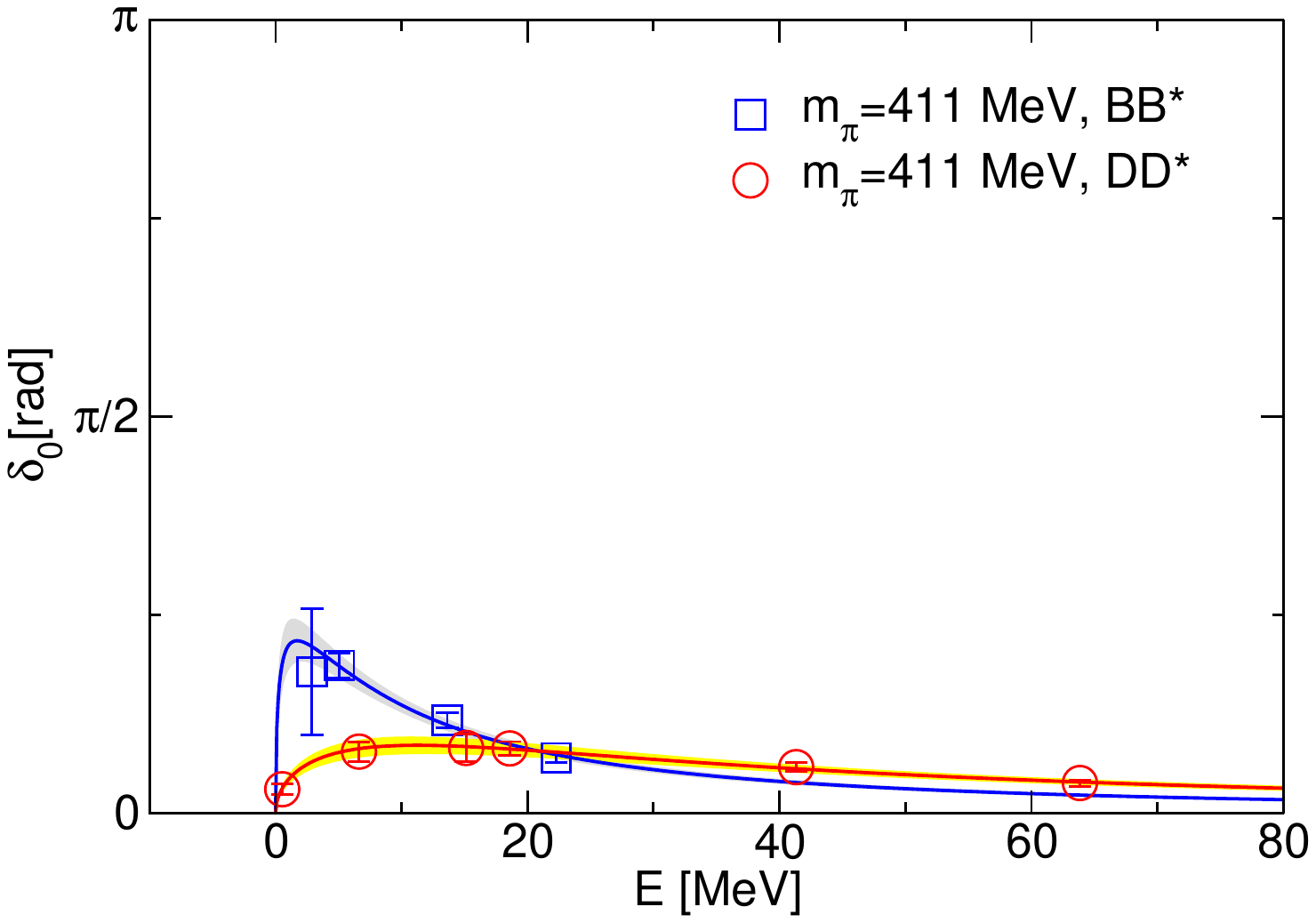} 
\caption{
$S$-wave scattering phase shifts for the $DD^*$ and  $BB^*$ scatterings
in the $I=0$ channel:
$m_\pi=295~\mathrm{MeV}$ (left) and 
$m_\pi=411~\mathrm{MeV}$ (right).
} 
\label{fig:plot_delta0}
\end{figure*}

\begin{figure*}
\centering
\includegraphics*[width=.48\textwidth,bb=0 0 792 612,clip]{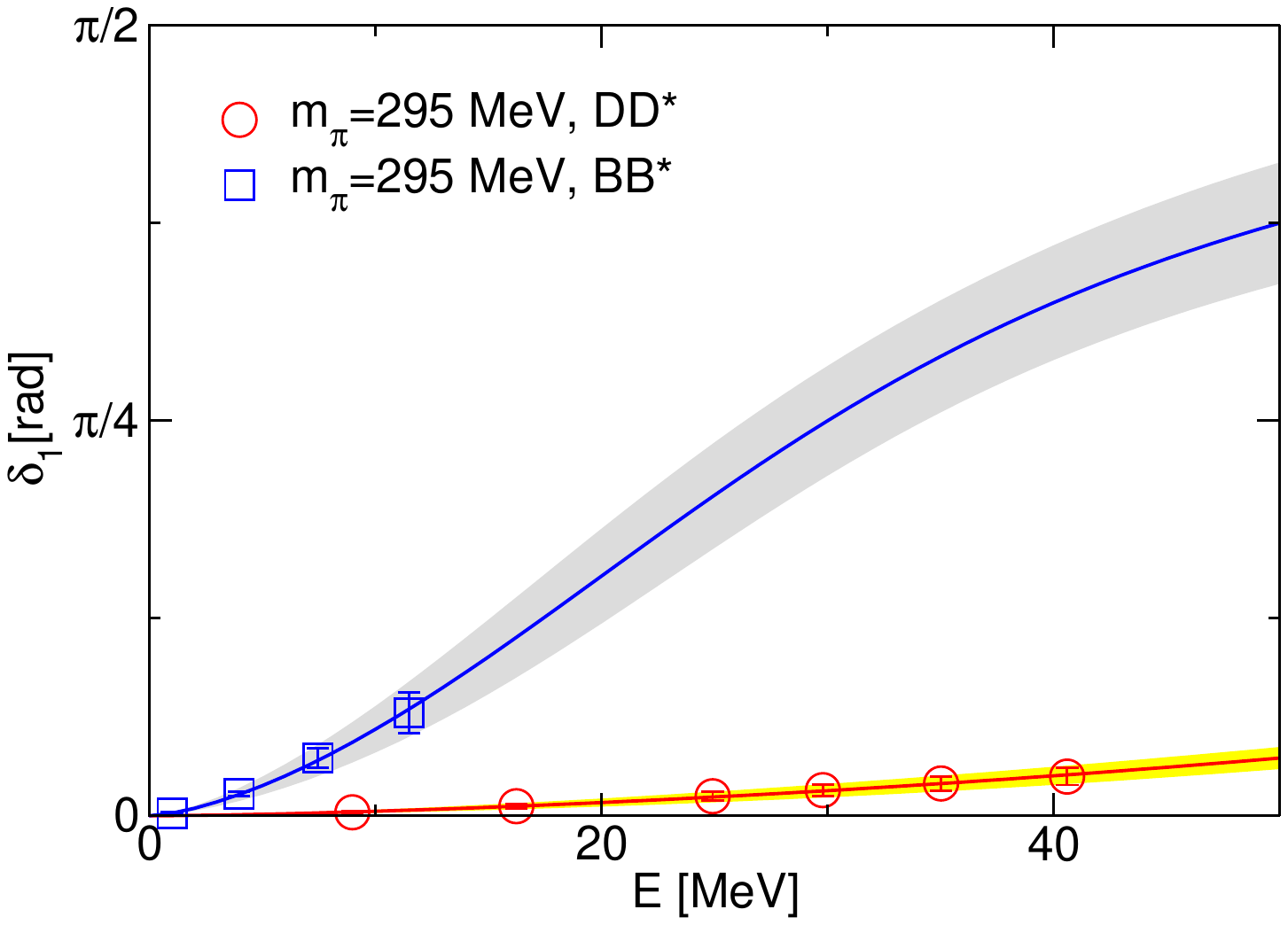} 
\includegraphics*[width=.48\textwidth,bb=0 0 792 612,clip]{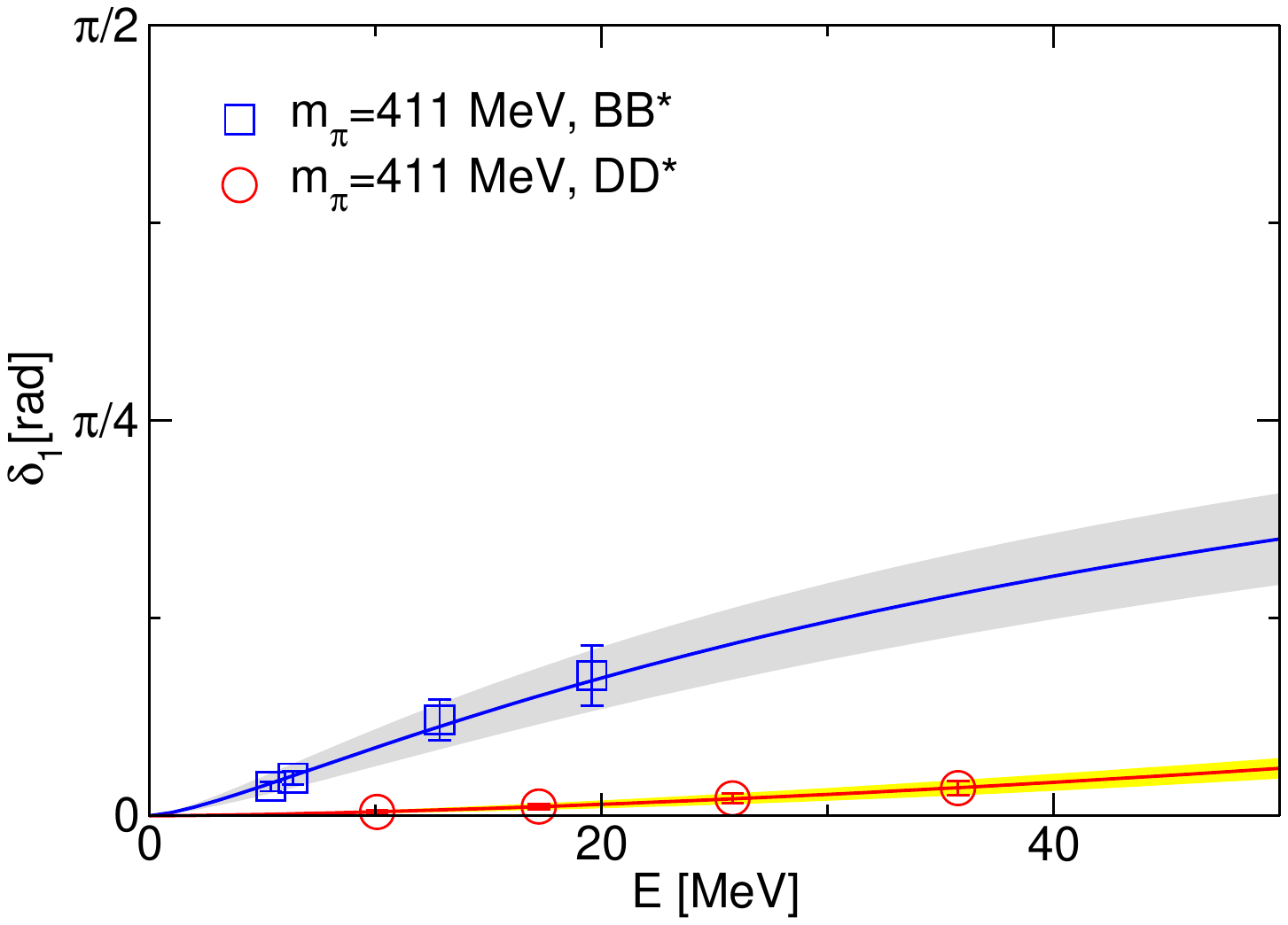} 
\caption{
$P$-wave scattering phase shifts for the $DD^*$ and $BB^*$ scatterings
in the $I=0$ channel:
$m_\pi=295~\mathrm{MeV}$ (left) and 
$m_\pi=411~\mathrm{MeV}$ (right).
}
\label{fig:plot_delta1}
\end{figure*}

The above procedure was carried out for four different combinations
of two light quarks ($m_\pi=295$ and 411 MeV) and two heavy-flavor quarks (charm and bottom). The results for $\delta_0$ and $\delta_1$ are shown 
as a function of the two-meson energy $E$ measured from the respective thresholds
in Fig.~\ref{fig:plot_delta0} and Fig~\ref{fig:plot_delta1}.

Both $S$-wave (Fig.~\ref{fig:plot_delta0}) and $P$-wave (Fig~\ref{fig:plot_delta1}) scattering phase shifts are positive, reflecting the attractive interaction
between the $D$ and $D^{*}$ ($B$ and $B^{*}$) states in the $I=0$ channel.
Although neither $T_{cc}$ nor $T_{bb}$ states could be observed as deeply bound states of $DD^{*}$ or $BB^{*}$ in our study, the weak attraction seen in both channels becomes stronger as $m_\pi$ decreases from 411 MeV to 295 MeV. 
Especially, for the $BB^{*}$ case, the unitary limit ($\lim_{k\rightarrow 0} k\cot \delta_0(k) \approx 0)$ is reached at $m_\pi~=$ 295 MeV, and the peculiar behavior of the scattering phase shift at $E~=$ 0 MeV suggests the formation of a shallow bound state~\footnote{The isospin projection was incorrect in our previous analysis where there was no sign of bound state formation.}. We can therefore expect a deeply bound state at lighter pion masses.

\begin{table*}[ht]
\begin{center}
%\begin{ruledtabular}
\begin{tabular}{cccccc}
\hline
Flavor & $\kappa_{h}$ & $\nu$ & $r_{s}$ & $c_{B}$ & $c_{E}$ \\ \hline
Charm   & $0.10819$ \ & $1.2153$ \ & $1.2131$ \ & $2.0268$ \ & $1.7911$  \\
Bottom  & $0.03989$ \ & $2.9570$ \ & $2.5807$ \ & $4.0559$ \ & $2.8357$  \\
\hline
\end{tabular}
\caption{
Parameters of the RHQ action for charm and bottom quarks used in this work.
The parameter set of the charm quark was determined in Ref.~\cite{Kawanai:2011jt}.
}
\label{Tab:QuarkParam}
%\end{ruledtabular}
\end{center}
\end{table*}
%

%
%\begin{table}[t]
\begin{table*}[ht]
\begin{center}
%\begin{ruledtabular}
\begin{tabular}{cccccc}
\hline
Ensemble&  $(\kappa_{ud}, \kappa_{s})$ & $M_{\pi}$ [MeV] & \# of configs.
\\
\hline
A & (0.13770, 0.13640) & 295(2) & 799 \\
B & (0.13754, 0.13640) & 411(1) & 450 \\
%C & (0.13727, 0.13640) & 569.3(8) & 400 \\
\hline
\end{tabular}
\caption{Simulation parameters of PACS-CS configurations~\cite{Aoki:2008sm}.}
\label{Tab:Param}
%\end{ruledtabular}
\end{center}
\end{table*}
%

%
%\begin{table}[t]
\begin{table*}[ht]
\begin{center}
%\begin{ruledtabular}
\begin{tabular}{cccccc}
\hline
Ensemble & $M_{D}$ [GeV] & $M_{D^*}$ [GeV] & $M_{B}$ [GeV] & $M_{B^*}$ [GeV]
\\
\hline
A & 1.877(2) & 2.030(4) & 5.258(3) & 5.310(4) \\
B & 1.901(2) & 2.056(4) & 5.291(5) & 5.344(7) \\
%C &  &  & \\
\hline
\end{tabular}
\caption{Mass spectrum of $D$, $D^*$, $B$ and $B^*$ mesons.}
\label{Tab:Spect}
%\end{ruledtabular}
\end{center}
\end{table*}

%\clearpage
%%%%%%
\section{Summary}
\label{sec:summary}
In this talk, we presented our results on the calculation of the $S$-wave and $P$-wave scattering phase shifts of the $DD^*$ and $BB^*$ systems in the $I=0$ channel using 2+1 flavor PACS-CS gauge ensembles simulated at $m_\pi=295$ and 411 $\mathrm{MeV}$.
The twisted boundary condition allows us to treat any small momentum
on the lattice through the variation of the twist angle, continuously.
Therefore, we can determine the low-energy scattering phase shifts near the threshold, where a rapid increase in scattering phase occurs as a precursor to bound state formation.
In our simulated pion mass region, we only observed an attractive interaction between $D$ and $D^{*}$ states in the $I=0$ channel, which was not strong enough to form bound states. However, for the case of $BB^*$, we observe that the unitary limit is reached at $m_\pi=295$ MeV, and the peculiar behavior of the scattering phase shift appears at $E~=$ 0 MeV. This suggests the formation
of a shallow bound state. We can therefore expect a deeply bound state
at lighter pion masses at least for the $BB^{*}$ system.

%--- acknowledgments ------------------------------------------------  
\begin{acknowledgments}
M.~N. is supported by Graduate Program on Physics for the Universe (GP-PU) of Tohoku University.
Numerical calculations in this work were partially performed using Yukawa-21 at the Yukawa Institute Computer Facility. This work was also supported in part by Grant-in-Aids for Scientific Research form the Ministry of Education, Culture, Sports, Science and Technology (No. 22K03612) and JSPS Research Fellows (No. 24KJ0412).
\end{acknowledgments}

\end{document}